\documentclass[final,5p,times,twocolumn,sort&compress]{article}
\usepackage[english]{babel}
\usepackage{amssymb}
\usepackage{color,graphicx}
\usepackage{amsmath}
\usepackage{amsthm}
\usepackage{mathtools}
\usepackage{amsbsy}
\usepackage{bbm}
\usepackage{bm}
\usepackage{multicol}
\usepackage{comment}
\usepackage{booktabs}
\usepackage{array}
\usepackage{hyperref}
\usepackage{enumerate}

\newcommand{\id}{\mathbb{I}}

\newcommand{\ra}{\rangle}
\newcommand{\la}{\langle}
\newcommand{\beq}{\begin{eqnarray}}
\newcommand{\eeq}{\end{eqnarray}}
\newcommand{\bra}[1]{\ensuremath{\langle #1 |}}
\newcommand{\ket}[1]{\ensuremath{| #1 \rangle}}

\newtheorem{theorem}{Theorem}[section]
\newtheorem{corollary}{Corollary}[theorem]
\newtheorem{proposition}{Proposition}[section]
\newtheorem{definition}{Definition}[section]
\newtheorem*{remark}{Remark}

\makeatletter
\newcommand{\proofpart}[2]{%
  \par
  \addvspace{\medskipamount}%
  \noindent\emph{Part #1: #2}\par\nobreak
  \addvspace{\smallskipamount}%
  \@afterheading
}
\makeatother

\usepackage[authoryear]{natbib}
\setcitestyle{square, comma, authoryear}
\begin{document}

\title{Proof of quantum-to-classical transition for the center of mass of quantum many-body systems}

\author{Marco Bilardello \footnote{email: marco.bilardello@gmail.com}}
\maketitle
\begin{abstract}
The classical limit of quantum mechanics is investigated, by focusing on the study of the center of mass of a many-body system where each particle is described by quantum mechanics. We study how, in the limit when the number of particles diverges and under quite general assumptions, the center of mass of the system is not anymore described by quantum mechanics but by classical mechanics: the center of mass of the system becomes with a well-defined position and momentum, the state of the center of mass is fully determined by its position and by its momentum, and its dynamics is given by the classical law of dynamics. In order to get this result, three assumptions on the many-body system are necessary: the total mass of the system must be much larger than the mass of each particle composing the system; at most a finite number of particles has non-zero correlation in position with an infinite number of particles; finally, when the number of particles composing the system diverge, the variance in position of each particle converges to a finite value. These assumptions are commonly realized for a macroscopic solid and liquid systems. The results obtained in this paper show how the classical limit of quantum mechanics can be naturally achieved without the need to modify quantum mechanics, without the need to have an external environment and without the need of changing the interpretations of quantum mechanics.
\end{abstract}




\section{Introduction}
Quantum mechanics is nowadays one of the most successful physics theory mankind has ever produced, allowing us to understand how microscopic systems work. Before quantum mechanics, classical mechanics was the main theory, that still works perfectly well for macroscopic systems but fails in describing microscopic systems. Despite the fact that quantum mechanics is considered to be a complete theory, it is still missing how to retrieve classical mechanics by starting from quantum mechanics when a limit to the macroscopic is performed. This has been an open problem since the very beginning of quantum mechanics, and over the decades several solutions have been proposed to solve this problem, from decoherence due to interaction with an external environment~\citep{joos2013decoherence} to stochastic modifications of the Schroedinger equation~\citep{ghirardi1986unified} and many more~\citep{bohm1952suggested},~\citep{everett1973theory}. One of the most interesting way of trying to achieve the quantum-to-classical transition has been by studying the limit where the number of particles of a many-body system goes to infinite. In this formalism, it has been proved that thermodynamical classical properties can be obtained as a limit of quantum mechanical ones~\citep{simon1980classical}, and it has been formalized the classical limit as $\hbar \to 0$~\citep{yaffe1982large}. However, none of these proposed solutions actually allows to obtain classical mechanics (which is properly described in terms of phase space and Poisson brakets) for systems composed by components which follows quantum mechanics, and there is even no agreement on what could be the cause of a system to be described by classical properties but not by quantum mechanics.

Goal of this paper is to provide an explanation on the quantum-to-classical transition of a many-body system, where each particle composing the system is describe by quantum mechanics but the macroscopic object follows classical mechanics.  We will show how, under very general assumptions, the algebra of the operators describing the position and the momentum of the center of mass of the system becomes abelian in the limit when the number of particle composing the system goes to $+\infty$, i.e. when the system became effectively macroscopic. This result immediately allows to express any pure state of the system by its (well-defined and localized) position of the center of mass and total momentum, and we derive the set of pure states composing the phase space of classical mechanics. We will see how, for macroscopic systems, is then impossible to be in a quantum superposition of states, and, also, we will retrieve the classical Hamiltonian equations of motions for the center of mass.

This article is structured as follows. In Sec.~\ref{sec:macro}, we formalize the limit from microscopic to macroscopic systems. In Sec.~\ref{sec:results}  we outline the main results in this paper showing how classical mechanics is achieved when moving to macroscopic properties of many body systems: the center of mass is described in terms of phase space and Hamiltonian dynamics. In Sec.~\ref{sec:proof} we outline the proof of the main theorems. Finally, in Sec.~\ref{sec:end} we discuss the physical meaning of the assumptions and we outline the final conclusions.
\section{Macroscopic limit of a many-body system}
\label{sec:macro}
We consider a system of interacting, massive particles. When the system has a finite $N+1$ number of particles, it is fully described by quantum mechanics, where the set of operators $\mathcal{A}_{N+1}$ is the Weyl algebra generated by the position operator $\hat{x}_i$ and momentum operator $\hat{p}_i$ for each particle $i = -\frac{N}{2}, \dots, \frac{N}{2}$, with commutation relation defined by the following relation:
\begin{equation}
\label{eq:weyl}
e^{i \alpha\hat{x}_j }e^{i \beta \hat{p}_k } = e^{- i \alpha \beta \hbar \delta_{j,k} } e^{i \beta \hat{p}_k } e^{i \alpha\hat{x}_j }.
\end{equation}
It is always possible to represent the bounded operators acting on the Hilbert space $\mathbb{H}_{\omega_{N+1}}$ either in terms of the particles operators $\hat{x}_i$, $\hat{p}_i$ or in terms of center of mass
\begin{equation}
\begin{aligned}
\hat{X}_{cm}^{N+1} &= \sum_{i = -\frac{N}{2}}^{\frac{N}{2}} \frac{m_i}{M} \hat{x}_i \\
\hat{P}_{cm}^{N+1} &= \sum_{i = -\frac{N}{2}}^{\frac{N}{2}} \hat{p}_i,
\end{aligned}
\end{equation}
where $M$ is the total mass of the system and each particle has mass $m_i$, and $N$ pairs of relative coordinates operators $\hat{r}_{k}^{N+1}$, $\hat{q}_{k}^{N+1}$, where $k = -\frac{N}{2}, \dots, -1, 1, \dots, \frac{N}{2}$ spans the relative coordinates (also known as Jacobi coordinates~\citep{amorim2019center},~\citep{fortunato2017diagonalization}). The information probed by these two sets of operators is different: the center of mass operators allow to get the average, collective information of the whole system, whereas relative coordinates operators allow to get information relative to internal behaviour of the system such as vibrational states for example. It is, however, the center of mass that represents the macroscopic properties of a system (as total mass, total momentum and localization in space), which are the main properties that generally participate in the interaction with external force and that we describe by using classical mechanics, e.g. for the dynamics of a solid object. 

The operator algebra $\mathcal{A}_{N+1}$ for can be written as joint union of two sub-algebras, $\mathcal{A}_{N+1} = \mathcal{A}_{cm}^{N+1} \cup \mathcal{A}_{rel}^{N+1}$. The Weyl relation~\ref{eq:weyl} still holds in each of the two sub-algebras
\begin{equation}
\label{eq:weyl_cm}
e^{i \alpha\hat{X}_{cm}^{N+1} }e^{i \beta \hat{P}_{cm}^{N+1} } = e^{- i \alpha \beta \hbar } e^{i \beta \hat{P}_{cm}^{N+1} } e^{i \alpha\hat{X}_{cm}^{N+1} }.
\end{equation}
\begin{equation}
\label{eq:weyl_rel}
e^{i \alpha\hat{r}_{k}^{N+1} }e^{i \beta \hat{q}_{l}^{N+1} } = e^{- i \alpha \beta \hbar \delta_{l,k} } e^{i \beta \hat{q}_{l}^{N+1} } e^{i \alpha\hat{r}_{k}^{N+1} }.
\end{equation}
Moreover, we have that $\forall A_{cm} \in \mathcal{A}_{cm}^{N+1}, B_{rel} \in \mathcal{A}_{rel}^{N+1}$, the commutator $[A_{cm}, B_{rel}] = 0$.

Dynamics is described in terms of one-parameter group of *-automorphisms $\alpha_{t}^{N+1}$ of $\mathcal{A}_{N+1}$, and it is generally generated by a Hamiltonian
\begin{align}
\label{hamN}
\hat{H}_{N+1} &= \sum_{i = -\frac{N}{2}}^{\frac{N}{2}} \left( \frac{\hat{p}_{i}^{2}}{2 m_i}   + V_{N}^{ext} (\hat{x}_i)\right) \\
&+ \sum_{i = -\frac{N}{2}}^{\frac{N}{2}}  \sum_{j \neq i} V_{N}^{int} (|\hat{x}_i - \hat{x}_j|) \nonumber \\
&= \hat{H}_{cm} + \hat{H}_{rel}. \nonumber
\end{align}
Here, we assume that the external potential allows to decouple the center of mass and internal degrees of freedom. This is the case that has generally interest in physics, and it is satisfied by common cases such as box potential or harmonic potential. 
States are defined as linear, positive and continuous functionals $\omega_{N+1}$ over the algebra $\mathcal{A}_{N+1}$.

To perform a macroscopic limit we have to compute a limit $N \to +\infty$. The limit operator algebra is generally the closure, under certain topology, of $\mathcal{A}_0 = \cup_N \mathcal{A}_{N}$. The norm-closure of $\mathcal{A}_0$ is generally indicated with $\mathcal{A}$, and it is the local-algebra of the system. The immediate question is whether it is possible to express $\mathcal{A}$ as union of norm-closure of center of mass and relative algebras, but this is not possible. It is indeed easy to check that the center of mass operators do not converge in norm topology, and a weaker topology needs to be defined. Indeed, if we take e.g. the total momentum operator $\hat{P}_{cm}^{N+1}$, we have that, if e.g. $M > N$:
\begin{equation}
|| e^{i x \hat{P}_{cm}^{M+1}} - e^{i x \hat{P}_{cm}^{N+1}} || = 2
\end{equation}
so it is not a Cauchy sequence. 

If we write $<\psi| A^{\dagger} e^{i x\hat{P}_{cm}^{M+1}} B e^{-i x\hat{P}_{cm}^{M+1}} |\psi> = \omega (A^{\dagger} \alpha_{M}^{x}(B))$, then the limit exists in weak topology:
\begin{align}
&|\omega(A^{\dagger} \alpha_{M}^{x} {B}) - \omega(A^{\dagger} \alpha_{N}^{x} {B})|^2 \nonumber \\
&\leq \omega(A^{\dagger} A) \omega( (\alpha_{M}^{x} {B} - \alpha_{N}^{x} {B})^{\dagger}(\alpha_{M}^{x} {B} - \alpha_{N}^{x} {B})). 
\end{align}
Ignoring the first, constant term, we can rewrite the second term and we get:
\begin{align}
&\omega( \alpha_{M}^{x} (B^{\dagger} B)) + \omega( \alpha_{N}^{x} (B^{\dagger} B)) - \nonumber \\
&\omega( \alpha_{M}^{x} (B^{\dagger}) \alpha_{N}^{x} (B)) - \omega( \alpha_{N}^{x} (B^{\dagger}) \alpha_{M}^{x} (B)).
\end{align}
By noticing that, for every $B \in \mathcal{A}, \, \epsilon >0$ there is $n \in \mathbb{N}$ such that exists $B_n \in \mathcal{A}_n$ such that $|| B_n - B|| < \epsilon$, we can rewrite the above relation and having terms like the followings
\begin{align}
&\omega( \alpha_{M}^{x} (B_{n}^{\dagger} B_{n})) + \omega( \alpha_{N}^{x} (B_{n}^{\dagger} B_{n})) - \nonumber \\
&\omega( \alpha_{M}^{x} (B_{n}^{\dagger}) \alpha_{N}^{x} (B_{n})) - \omega( \alpha_{N}^{x} (B_{n}^{\dagger}) \alpha_{M}^{x} (B_{n})) \nonumber \\
& \to 0
\end{align}
because if $M,N$ are large enough, they will be both larger than $n$. The remaining terms are all bounded from above by the norm $||B - B_n|| < \epsilon$.

As a consequence, we will need to work on a wider algebra $\mathcal{A}^{\prime \prime}$, which is the universal Von Neumann algebra of $\mathcal{A}$. The convergence of the states $\omega_{N+1}$ generally occurs in the weak topology of the dual $\mathcal{A}^{\prime}$ of $\mathcal{A}$.

In order to be sure that, in the limit $N \to +\infty$, sub-algebras of center of mass and relative coordinates, and the dynamics $\alpha_{t}^{N+1}$, converge properly, we need to define a set of states $F$ where, from the physics point of view, we expect that these limits exists and where we expect to achieve the usual classical mechanics. The main result of this paper is that, under the topology $\tau_F$ on $\mathcal{A}^{\prime \prime}$ defined from the set of states F, these limits actually exists and we actually obtain the classical mechanics (in terms of states, operators and dynamics) for the center of mass of the system.
\begin{definition}
We say that $\omega \in F$ if it satisfies the following conditions:
\begin{enumerate}
\label{F_states}
\item \label{assumption0} $\omega$ is composed by an infinite number of particles. This is generally obtained by having a sequence of states over an increasing number of particles $\omega_N$, with $\omega = \lim_{N \to + \infty} \omega_N$ in weak topology of the dual space $\mathcal{A}^{\prime}$;
\item \label{assumption1} in the limit above, the ratio of the mass of any particle $m_i$ over the total mass $M$ goes to 0, i.e. $\frac{m_i}{M} \xrightarrow{\tiny \text{ \tiny N} \to + \infty} 0$;
\item \label{assumption2} as the total number of particles diverges, there is at most only a finite number of particles which shows non-zero space correlation with an infinite number of particles. This means that quantities such as $|\la \hat{x}_i \hat{x}_j \ra -\la \hat{x}_i \ra \la \hat{x}_j \ra|$ (and similarly for higher order of correlation) are non zero for every particle $j$ only for a finite number of particles $i$; 
\item \label{assumption3} as the total number of particles diverges,  the quantities $|\la \hat{x}_{i}^{k} \ra -\la \hat{x}_i \ra^{k} |$ converge to a finite value $\forall k \geq 2$,
\end{enumerate}
\end{definition}
In the next section, we are going to discuss about the definition above. Here, we want to point out that, if $\omega \in F$, then also 
\begin{align}
\label{coherentstates}
&\omega(x,p) (A) = \lim_N \omega_{N} (x,p) (A) \notag \\
&=\lim_N \bra{\psi_{N}} e^{i \frac{x \hat{P}_{cm}^{N}}{\hbar}} e^{i \frac{p \hat{X}_{cm}^{N}}{\hbar}} \hat{A} \notag \\
& e^{- i \frac{p \hat{X}_{cm}^{N}}{\hbar}} e^{- i \frac{x \hat{P}_{cm}^{N}}{\hbar}} \ket{\psi_{N}}, 
\end{align}
is an element of $F$, where $\ket{\psi_{N}}$ is the cyclic vector in the representation of $\omega_N$. Of particular importance in physics are two cases: where $\omega$ is the minimal energy state (ground state), and where $\omega$ is the thermal state defined as a Gibbs convex sum of energy eigenstates. In general, in physics every possible state that a system can achieve can be expressed as a modification of a minimum-energy state
 (ground state) if the system is not interacting with a second system (external environment), or as a modification of a thermal equilibrium that the system reached after interacting with the external environment. These modifications are generally local, i.e. expressed as operators in the quasi-local algebra $\mathcal{A}$, while the global properties of the system are expressed by operators in the Von Neumann algebra $\mathcal{A}^{\prime \prime}$.
The states~\ref{coherentstates} have also well-precise meaning: it is the collection of states where internal (relative) variables are in equilibrium, and the center of mass variables are translated in space by an amount $x$ and boosted in momentum by an amount $p$.

It has been proved~\citep{morchio1987mathematical} that, to have convergence of states and dynamics in the weak topology $\tau_F$ in the Von Neumann algebra $\mathcal{A}^{\prime \prime}$, set of states F need to have the following properties:
\begin{enumerate}[i]
\label{F_conditions}
\item closed under linear combinations; 
\item norm-closed and separating, i.e. $\omega(A) = 0 \, \forall \omega \in F \implies A =0$; 
\item it is stable under local operations, i.e. $\omega \in F \implies \forall A,B \in \mathcal{A}, \, \omega(A \cdot B) \in F$. 
\end{enumerate}
Let's call $\mathcal{M}$ to be the closure of $\mathcal{A}$ with $\tau_F$ topology. 
Indeed, the following propositions can be proved:
\begin{proposition}
\label{prop1}
Given F set of states defined as above over an algebra $\mathcal{A}$, then an element of the dual  $\mathcal{A}^{\prime}$ is in F iff it is $\tau_F$ continuous.
\end{proposition}
\begin{proposition}
\label{prop2}
The limit
\begin{equation}
\label{limit_dynamics}
\alpha^t (A) = \lim_{N \to + \infty} \alpha_{N}^{t} (A),
\end{equation}
exists in $\mathcal{M}$ for any $A \in \mathcal{A}$, and it is $\tau_F$ continuous on $\mathcal{A} \iff  \lim_{N \to + \infty} \alpha_{N}^{t^*} \omega$ exists for any $\omega \in F$ in the weak* topology induced by $\mathcal{A}$ on its dual, and it belongs to F.

Under such conditions, the mapping defined on $\mathcal{A}$ by~\ref{limit_dynamics} has a unique continuous extension to $\mathcal{M}$, preserving sums, multiplication by scalar and *. This extension is obtained by transposing the limit $\alpha^{t^*} = \lim_{N \to + \infty} \alpha_{N}^{t^*}$ on the dual of $\mathcal{A}$.
\end{proposition}
\begin{proposition}
\label{prop3}
The mapping defined in~\ref{limit_dynamics} is a morphism of $\mathcal{M}$ iff $\alpha^t$ converges on $\mathcal{A}$ in the ultrastrong topology defined by F on $\mathcal{M}$.

Propositions~\ref{prop2} and~\ref{prop3} are authomatically satisfied if any of these conditions is realised:
\begin{enumerate}
\item $\alpha_{N}^{t}$ converges ultrastrongly on $\mathcal{A}$ and weakly on $\mathcal{M}$;
\item $\alpha_{N}^{t}$ converges in the norm topology on the set of states F;
\item there exists a C*-subalgebra $\mathcal{B}$ of $\mathcal{M}$, $\mathcal{A} \subset \mathcal{B}$, such that, for any $B \in \mathcal{B}$, the weak limit of $\alpha_{N}^{t} (B)$ exists uniformly on F, and it is weakly continuous.
\end{enumerate}
\end{proposition}

Proof of~\ref{prop1},~\ref{prop2},~\ref{prop3} can be found in~\citep{morchio1987mathematical}, propositions 2.1, 2.2 and 2.3.

We have to prove that the set of states defined in~\ref{F_states} satisfies the conditions~\ref{F_conditions}. This is proved in the following theorem:
\begin{theorem}
\label{FTheorem}
The set of states~\hyperref[F_states]{F} satisfies the conditions for proposition~\ref{prop1}.
\end{theorem}
\begin{proof}
It is trivial to prove the point i). Let's prove point iii): given $A, B \in \mathcal{A}$, and $\omega \in F$, let's define $\omega_{AB} \equiv \omega (A \cdot B)$. Then, we have to prove that $\omega_{AB}$ satistifes conditions~\ref{assumption2},~\ref{assumption3}. But that is a direct consequence of $|\omega_{AB} (O)| \leq ||A|| ||B|| |\omega(O)|$.

Let's finally prove point ii). Let's assume that, for a given $\omega \in F$ and $A \in \mathcal{A}$, $||A|| \neq 0$, $\omega(A) = 0$. Let's prove that there exists at least one state $\nu \in F$ such that $\nu(A) \neq 0$. Since $A \in \mathcal{A}$, $\forall \epsilon > 0$ there is $\bar{N} \in \mathbb{N}$ such that exists $A_{\bar{N}} \in \mathcal{A}_{\bar{N}}$ such that $|| A - A_{\bar{N}}|| < \epsilon$. We can define a finite-particle state $\mu_{\bar{N}}$ such that $\mu_{\bar{N}}(A_{\bar{N}}) \neq 0$. Then, the state $\nu$ such that:
\begin{equation}
\nu(O) = \frac{1}{2}(\mu_{\bar{N}} (O_{\bar{N}}) + \omega(O))
\end{equation}
has $\nu(A) \neq 0$ and it is clearly in F.
\end{proof}

Assumption~\ref{assumption2} can be achieved when there is no long distance correlation among particles, i.e. for every operators $O_{i},O_{j} \in \mathcal{A}$ localised, respectively, in the regions $\Lambda_i, \Lambda_j \subset \mathbb{R}^3$, $\Lambda_i \cap \Lambda_j = \emptyset$, then
\begin{equation}
\label{assumption2'}
\lim_{|i-j| \to +\infty} |\omega(O_i O_j) - \omega(O_i) \omega(O_j)| = 0 \tag{II'}.
\end{equation}
This is not an unusual situation, and it is actually realised everytime a system is macroscopic and there is a limit of number of particles to infinite, like in statistical mechanics with the usual thermodynamical limit (see, e.g., ~\citep{georgii2011gibbs} chapter 7). Assumption~\ref{assumption3}, on the other hand, specifies that the variance of the position of each particle cannot diverge. To achieve this, two things need to realize: 
\begin{enumerate}
\item \label{assumption_particle_interaction} the interparticle interaction $V_{N}^{int} (|\hat{x}_i - \hat{x}_j|)$ has to be attractive and strong enough to determine that the all but a finite number of particles stay in a bound state, i.e. constraining the particles to occupy a finite region of space around the center of mass;
\item \label{assumption_ext_interaction} there is an external potential which traps the center of mass of the system in a finite region of space.
\end{enumerate}


We notice, however, that the case where there is no external potential is not included among the hypothesis of~\ref{mainTh}. This is a case we still can consider, by slightly modifying the set of states $F$ into $F^{\prime}$:
\begin{definition}
\label{F_prime}
We say that $\omega \in F^{\prime}$ if conditions~\ref{assumption0},~\ref{assumption1},~\ref{assumption2} are preserved, but where condition~\ref{assumption3} is replaced by the following conditions:
\begin{enumerate}
\item the interparticle interaction $V_{N}^{int} (|\hat{x}_i - \hat{x}_j|)$ has to be attractive and strong enough to determine that the all but a finite number of particles stay in a bound state, i.e. constraining the particles to occupy a finite region of space around the center of mass;
\item \label{assumption_no_trap} the system is symmetric under space translation, i.e. it is invariant under transformation generated by $\hat{P}_{cm}$.
\end{enumerate}
\end{definition}
\begin{theorem}
\label{F_prime_Theorem}
The set of states ~\hyperref[F_prime]{$F^{\prime}$} satisfies the conditions for proposition~\ref{prop1}.
\end{theorem}
\begin{proof}
i) is trivial to prove, and proving ii) goes as for the state F. The only thing to prove is iii), i.e. that if $\omega$ is invariant under space translations then $\omega_{AB} \equiv \omega(A \cdot B)$ is invariant as well, for any $A, B \in \mathcal{A}$. But this holds, because A, B are quasi-local operators, i.e. can be approximated as operators involving a finite number of particles, which cannot modify global properties. Indeed,
\begin{align}
&\omega_N( e^{- i \alpha \sum_{i = -J}^{J} \frac{m_i}{M} \hat{x}_i} O(\hat{X}_{cm}^{N},\hat{P}_{cm}^{N}) e^{i \alpha \sum_{i =-J}^{J} \frac{m_i}{M} \hat{x}_i} )\\
&= \omega_N(O(\hat{X}_{cm}^{N},\hat{P}_{cm}^{N} +\hbar\alpha \sum_{i = -J}^{J} \frac{m_i}{M})) \nonumber\\
&\to \omega(O(\hat{X}_{cm},\hat{P}_{cm})). \nonumber
\end{align}
as of consequence of~\ref{assumption1}.
\end{proof}
Finally, the mascroscopic limit needs to have a finite total mass M of the system. The idea from physics is that, when having macroscopic operators (i.e. involving an infinite number of particles) any information on microscopic properties (i.e. involving a finite number of particles) is lost. We need to properly rescale the masses $m_i$ of the particles in order to have the total mass $M = \sum_i m_i$ is finite also in the limit of $N \to +\infty$. The simplest way is then to rescale the masses of each particle,
\begin{equation}
\label{massrescale}
m_i \to m_{i}^{N+1} = \frac{m_i}{N+1}
\end{equation}
and having that
\begin{equation}
\label{massconv}
\lim_{N} \frac{1}{N+1} \sum_{i = -\frac{N}{2}}^{\frac{N}{2}} \frac{m_i}{M} = 1.
\end{equation}
In statistical mechanics, in general the total mass M diverges while the particles masses $m_i$ are constant: this because the meaning of the limit is different. Indeed, keeping the masses $m_i$ constant while M diverges means that you are still probing the microscopic properties of the system, and the values of the macroscopic properties are, in comparison, infinite. We reverse this situation here: we aim to probe the macroscopic properties of the system and, in comparison, the values of the microscopic properties of the system are negligible.

\section{Main results}
\label{sec:results}
The first main result of the paper is expressed by the following theorem:
\begin{theorem}
\label{mainTh}
Given a system of $N$ particles where the algebra of observables is $\mathcal{A}_N$, and let's assume that the cyclic states $\omega_N$ converge weakly to $\omega$ in the limit $N \to + \infty$.

If $\omega \in F$, then the GNS representation of the sub-algebra of the center of mass $\pi_{\omega_N} (\mathcal{A}_{cm}^{N})$ of the system converges strongly to an abelian sub-algebra of the universal Von Neumann algebra $\mathcal{M}$. The macroscopic properties of $\omega$ are then determined by well-defined and localized values of the position and momentum of the center of mass. 
\end{theorem}
If we assume that, in general, the system is symmetric under space translation, we can however still prove that, under the same macroscopic limit, the center of mass becames fully classical. This is a direct consequence of the following theorem:
\begin{theorem}
\label{mainTh3}
If, in the macroscopic limit, either the center of mass position $\hat{X}_{cm}$ or the total momentum $\hat{P}_{cm}$ converges to a multiple of the identity, then the sub-algebra $\mathcal{A}_{cm}^{N}$ converges strongly to an abelian algebra, and the macroscopic properties of the extremal ground state $\omega$ are determined by well-defined and localized values of the position and momentum of the center of mass.
\end{theorem}
\begin{corollary}
\label{converge_no_trap}
Given $\omega \in F^{\prime}$, then the sub-algebra $\mathcal{A}_{cm}^{N}$ converges strongly to an abelian algebra and in extremal ground states, macroscopic properties behave classically.
\end{corollary}
\begin{proof}
This is a direct consequence of~\ref{mainTh3} after imposing that the ground states need to be translationally invariant, with $\hat{P}_{cm}$ equal to a multiple of identity to minimize the energy.
\end{proof}

The macroscopic states that realize either~\ref{mainTh} or~\ref{mainTh3} are then many-body states where the particles interact with each other and the interaction need to have an attractive component. Examples of these systems are the usual solids and liquids, but not dilute gases, where quantum properties might still be present at the macroscopic level (e.g. in the case of Bose-Einsten condensation).

Let's focus now on the meaning of theorems~\ref{mainTh},~\ref{mainTh3}. If their assumptions hold, then a macroscopic system has always localized values for its position and its momentum. From these theorems, it is clear that classicality is achieved if and only if the cyclic state $\omega$ satisfies the appropriate assumptions. As a consequence, if $\omega$ satisfies~\ref{mainTh},~\ref{mainTh3}, then any state $\omega(x,p)$~\ref{coherentstates} defined as the state $\omega$ with center of mass shifted by an amount $x$ and with total momentum boosted by $p$, satisfies the same theorems.

If previous theorems define the static classical properties of the center of mass, we still need to discuss the dynamical properties. Classical dynamics can be achieved by adding one more assumption to those we previously added so far: dynamics does not have to create instantaneously infinite-range correlations. This can be achieved by assuming that correlation among distant particles are bounded by a finite term at any time $t>0$. This has been proved for several discret and continuous systems, and it is generally due to bounds (known as Lieb-Robinson bounds) on the spread of the correlations among localised operators~\citep{bravyi2006lieb},~\citep{nachtergaele2006lieb},~\citep{hinrichs2025lieb}:
\begin{equation}
\label{lieb_robinson}
|| [\alpha^{t} (A), B] || \leq C \exp ( - d(A,B) + v(\hat{V}_{\text{int}})t)
\end{equation}
for any $A,B \in \mathcal{A}$, where the $d(A,B)$ refers to a distance between the localised supports of operators $A,B$, and where the speed of correlation spread $v$ is generally function of the interaction among particles. Without going into the details, the important thing for this paper is that the above bound allows us to say that, for any finite particle operator $A_n$ and for any time $t>0$, there exists $k < + \infty$ such that $\alpha^t (A_n) \in \mathcal{A}_k$.

We can then move to the next theorem of this paper:
\begin{theorem}
\label{mainTh2}
If $\omega \in F$ (or $\omega \in F^{\prime}$), the same conclusions of~\ref{mainTh} can be reached when replacing the state $\omega$ by any state $\omega(x,p)$. This set of states then composes the classical phase space, and each macroscopic state is fully determined by the points $(x,p)$. Let's introduce $\hat{A}_N \in \mathcal{A}_{cm}^{N}$, where the sequence $\left\{ \hat{A}_N \right \}_{N \in \mathcal{N}} \to \hat{A} \in  \mathcal{A}_{cm}$ as $N \to + \infty$ in the strong-operator topology. Let's introduce the finite-particle dynamics $\alpha_{t}^{N} :  \mathcal{A}_N \to \mathcal{A}_N$ with the usual dynamics given by:
\begin{equation}
\alpha_{t}^{N} (\hat{A}_N) = e^{\frac{i}{\hbar} \hat{H}_N t} \hat{A}_N e^{\frac{-i}{\hbar} \hat{H}_N t}.
\end{equation}

In the topology defined by $F$ (or $F^{\prime}$), if the finite-particles dynamics $\alpha_{N}^{t}$ and the limiting dynamics $\alpha^{t}$ satisfy a Lieb-Robinson bound of the form~\ref{lieb_robinson}, then the map $\alpha^t$ is a *-automorphism in $\mathcal{M}$ and, when applied over a center of mass operator, the resulting dynamics is classical:
\begin{equation}
\frac{d}{dt} \omega(\alpha_t(A)) = \lim_N \frac{-i}{\hbar}\omega_N ([\hat{A}_N, \hat{H}_N]).
\end{equation}
The dynamics links the states $\omega(x,p)$ which share all the same energy level $E = \frac{p^2}{2 M} + V_{ext} (x)$ in a continuous trajectory of motion.
\end{theorem}
This last theorem completes the picture of how the quantum-to-classical transition occurs, by finalizing the construction of the classical phase space, and by retrieving the classical Hamiltonian dynamics for a point in the phase space. If the assumptions~\ref{assumption1},~\ref{assumption2} and~\ref{assumption3} hold, then the macroscopic dynamics maps localized, classical states into new localized, classical states, and the classical dynamics is achieved as macroscopic limit of the quantum dynamics.
\section{Proof of main results}
\label{sec:proof}
\begin{proof}[Proof of~\ref{mainTh}]
It is better to move step by step.
\proofpart{1}{Localization of position of the center of mass in the ground state}

The proof of the localization of the position of the center of mass in the ground state goes as follows. We will prove that, for any $n \in \mathbb{N}$, in the limit $N \to + \infty$ there is $|\omega_{N+1} \left ( \left( \hat{X}_{cm}^{N+1}\right)^n \right) - \omega_{N+1} \left (\hat{X}_{cm}^{N+1} \right)^n| \to 0$. Indeed,
\begin{align}
\label{localizedposition}
&\left|\omega_{N+1} \left ( \left( \hat{X}_{cm}^{N+1}\right)^n \right) - \omega_{N+1} \left (\hat{X}_{cm}^{N+1} \right)^n \right| = \nonumber \\
&\left|\frac{1}{(N+1)^n} \sum_{|j_1| \leq \frac{N}{2}}\dots \sum_{|j_n| \leq \frac{N}{2}} \frac{\prod_{l=1}^{n} m_{j_l}}{M^n} \right. \nonumber \\
&\left. \left( \bra{\psi_{N+1}} \prod_{l=1}^{n} \hat{x}_{j_l} \ket{\psi_{N+1}} - \right. \right. \nonumber \\
&\left. \left. \prod_{l=1}^{n} \bra{\psi_{N+1}} \hat{x}_{j_l} \ket{\psi_{N+1}}\right)\right| 
\end{align}
where we just expanded the center of mass position operator over its particle components. By looking at the terms in~\ref{localizedposition}, we notice that, by assumption~\ref{assumption2} the terms where all the indices $j_1, \dots, j_n$ are different with each other is equal to 0 except for a finite number of cases where the correlation does not diverge in the macroscopic limit (as a consequence of ~\ref{assumption3}), and then go to $0$ due to the factor $\frac{1}{(N+1)^n}$ in front of everything (coming from~\ref{massrescale}). With the same logic we can prove that also the cases where only one pair of indices $j_1, \dots, j_n$ is equal with each other, the terms go to $0$ in the macroscopic limit due to the fact that there are $(N+1)^{(n-1)}$ finite terms in the sums, which are again pushed to $0$ by the multiplicative factor $\frac{1}{(N+1)^n}$.
This of course works for all the remaining terms of the sums, proving the localization of the position for the ground state.

\proofpart{2}{Localization of position of the center of mass for every other state}
Given $\omega_{N+1}$, for any $k < +\infty$, and for any $\hat{A}, \hat{B} \in \mathcal{A}$ we have that:
\begin{align}
\label{macromicro}
&\omega_{N+1} \left( \hat{A} \left[\hat{X}_{cm}^{k+1},  \hat{P}_{cm}^{N+1}\right]\hat{B}\right) = \\
&\omega_{N+1} \left( \hat{A}\hat{B}\right) \left(\frac{-i\hbar}{N+1}\sum_{j = -\frac{k}{2}}^{\frac{k}{2}} \frac{m_j}{M}\right) \nonumber \\
& \to 0, \nonumber
\end{align}
because $|\omega \left( \hat{A}\hat{B}\right)| < +\infty$.

By using~\ref{macromicro}, it is easy to see that, for any $n, l, k < +\infty$, for any $\hat{A_n} \in \mathcal{A}_{n}$,$\hat{B_l} \in \mathcal{A}_{l}$, there is
\begin{align}
\label{localizedposition3}
&\lim_{N\to +\infty} \left |\omega_{N+1} \left (\hat{A_n} \left (\hat{X}_{cm}^{N+1} \right )^k \hat{B_l} \right ) - \right. \\
&\left. \left(\omega_{N+1} \left (\hat{X}_{cm}^{N+1} \right ) \right)^k \right| =0. \nonumber
\end{align}
Proof goes as for~\ref{localizedposition}, by taking a finite $N_0$ such that $\mathcal{A}_k \cup \mathcal{A}_l \subseteq \mathcal{A}_{N_0}$, and by splitting the first $N_0$ particles from the remaining particles. The term with $N_0$ then goes to 0 because of~\ref{massrescale} and the remaining term goes to 0 as seen in previous part of the proof. As a consequence, we have that, for any operator $F(\hat{X}_{cm}^{N+1})$ in the GNS representation $\pi_{\omega}$ the following limit holds: $\pi_{\omega}\left(F(\hat{X}_{cm}^{N+1}) \right) \to \omega\left(F(\hat{X}_{cm}) \right) \id$.

\proofpart{3}{The macroscopic algebra of the center of mass is abelian}

This is a direct consequence of~\ref{converge_no_trap}.
The convergence of the center of mass algebra to an abelian algebra implies that it is equivalent to the algebra of continuous function from the space of the spectrum of the generators of the algebra to the complex field. Moreover, the pure states are product states, i.e. they satisfy the relation $\omega(A B) = \omega(A) \omega(B)$ for any $A,B \in \mathcal{A}_{cm}$. In other words, the pure states are those which have well-defined position and momentum~\citep{bratteli2012operator}.
\end{proof}
\begin{proof}[Proof of~\ref{mainTh3}]
Let's suppose we have that, in the macroscopic limit, $\hat{P}_{cm}$ converges to a multiple of the identity. Then, it is enough to see that any extremal state has $\hat{X}_{cm}$ as a multiple of identity too. Let's assume that the state $\omega$ has $\omega(\hat{X}_{cm}) = 0, \omega(\hat{X}_{cm}^{2}) \neq 0$. Then, we have the operator $e^{i \hat{X}_{cm}}$ which is in the commutant but it is not a multiple of the identity. Then $\omega$ is not extremal.
\end{proof}
\begin{proof}[Proof of~\ref{mainTh2}]
Starting from $\omega_N (x,p) (A_N)$ (as introduced in~\ref{coherentstates}) for an operator of the Weyl algebra $A_N (X_{cm}^{N}, P_{cm}^{N})$, we can easily check its relation to the ground state $\omega_N$:
\begin{align}
&\omega_N (x,p) (A_N) = \omega_N (x,p) (A_N(X_{cm}^{N}, P_{cm}^{N})) \nonumber \\
& = \omega_N (A_N ((X_{cm}^{N} + x, P_{cm}^{N} + p)),
\end{align}
and this is enough to prove that~\ref{mainTh} applies also to the set of coherent states~\ref{coherentstates}. Each coherent state localizes then in position in $x + \omega(X_{cm})$ and localizes in momentum in $p + \omega(P_{cm})$.
In order to see that the conditions in the propositions~\ref{prop2},~\ref{prop3} realise, we will see that the dynamics converges in norm topology over $\mathcal{A}$ and that it converges in weak* topology in the dual $\mathcal{A}^{\prime}$ and that the state $\omega_t$ is still in $F$ (or $F^{\prime}$).
Firstly, let's see that, for any $A \in \mathcal{A}$, for any $\epsilon >0$ there is a finite-particle operator $A_n$ such that $|| A - A_n|| <\epsilon$, due to the norm-closure of the quasi-local algebra. Secondly, by using the bounds coming from Lieb-Robinson relation~\ref{lieb_robinson} we have that, after a fixed time $t>0$, $\alpha^t (A_n) \in \mathcal{A}_k$ for a certain finite $k$.
We then have:
\begin{align}
&|| \alpha_{N}^{t}(A) - \alpha^{t}(A)|| < || \alpha_{N}^{t}(A) - \alpha_{N}^{t}(A_n)|| +\nonumber \\ 
& || \alpha_{N}^{t}(A_n) - \alpha^{t}(A)|| \nonumber \\
& < || \alpha_{N}^{t}(A) - \alpha_{N}^{t}(A_n)|| + || \alpha^{t}(A) - \alpha^{t}(A_n)||  +\nonumber \\
& || \alpha_{N}^{t}(A_n) - \alpha^{t}(A_n)|| \nonumber \\
& < 3 \epsilon
\end{align}
where the first two $\epsilon$ are coming from the first two norms due to the norm-preserving properties of $\alpha_{N}^{t}, \alpha^{t}$ (in the norm-closed $\mathcal{A}$ we know that the limit maps $\alpha^t$ is at least a *-homomorphism), while, for the last norm, due to the Lieb-Robinson bounds there is a finite $N$ large enough such that $ || \alpha_{N}^{t}(A_n) - \alpha^{t}(A_n)|| < \epsilon$.

Since it converges in norm topology in $\mathcal{A}$, then it converges in weak-* topology in the dual space $\mathcal{A}^{\prime}$. If $\omega$ has no long-range correlation, then at any time $t>0$ the Lieb-Robinson bounds guarantee that there are no long-range correlations. Finally, if $\omega \in F^{\prime}$, then $\omega_t$ is still invariant under space translation. Indeed, given $\beta^x$ the *-automorphism determining space translation by $x$, we have that, for any $A \in \mathcal{A}$:
\begin{align}
&\beta^x \alpha^t (A) = \beta^x ( \alpha_{N}^{t} (A) +  \alpha^t (A) - \alpha_{N}^{t} (A)) \nonumber \\
&= \alpha_{N}^{t} \beta^x (A) + \beta^x ( \alpha^t (A) - \alpha_{N}^{t} (A)) \nonumber \\
& \to 0.
\end{align}  

Since the dynamics splits between center of mass and relative coordinates, then we can identify the dynamics $\alpha_{cm}^{t}$ as the *-automorphism on the subalgebra $\mathcal{A}_{cm}$ of $\mathcal{M}$. The equations of motion are then easily computed:
\begin{align}
\label{dynamicsposition}
&\frac{d}{dt} \omega_{N+1} \left(\hat{A} \, \alpha_{N+1}^{t}\left(\hat{X}_{cm}^{N+1} \right) \hat{B}\right) = \\
&\omega_{N+1}\left(\hat{A} \left[\hat{X}_{cm}^{N+1}, \frac{\left(\hat{P}_{cm}^{N+1} \right)}{2 M_{N+1}} \right]\hat{B}\right) \nonumber \\
& \to \frac{1}{M} \omega \left(\hat{P}_{cm} \right) \nonumber
\end{align}
where we defined $M_{N+1} = \frac{1}{N+1} \sum_{i = -\frac{N}{2}}^{\frac{N}{2}} m_j$.

Regarding the dynamics of the total momentum, we can compute it in a similar way:
\begin{align}
\label{dynamicmomentum}
&\frac{d}{dt} \omega_{N+1} \left(\hat{A} \, \alpha^{t}_{N+1}\left(\hat{P}_{cm}^{N+1} \right) \hat{B}\right) = \\
&\omega_{N+1}\left(\hat{A} \left[\sum_{j = - \frac{N}{2}}^{\frac{N}{2}} \hat{p}_j, \sum_{k = - \frac{N}{2}}^{\frac{N}{2}} V_{ext} (\hat{x}_k)\right]\hat{B}\right) \nonumber \\
&= \omega_{N+1}\left(\hat{A} \left[\hat{P}_{cm}^{N+1}, V_{ext} (\hat{X}_{cm}^{N+1})\right]\hat{B}\right) \nonumber \\
& \to - \omega \left(\frac{dV_{ext}}{dx} (\hat{X}_{cm})\right). \nonumber
\end{align}

As a consequence, in the macroscopic limit the pure states are not stationary, differently from the microscopic energy eigenstates. Stationary energy state can be obtained in the macroscopic limit but, generally, only as mixed states or in the case if there is only one pure state for a given energy level. Indeed, it is possible to easily check that the set of all the pure states $\omega(x,p)$ that have a fixed energy $E = \omega(x,p) (\hat{H}_{cm})$, i.e. such that:
\begin{equation}
\frac{p^2}{2M} + V_{ext} (x) = E
\end{equation}
are connected in a continuous trajectory solution of the classical Hamiltonian dynamics~\ref{dynamicsposition},~\ref{dynamicmomentum}. 
\end{proof}

\begin{remark}
For example, if the external potential is harmonic with frequency $\nu$, the limit dynamics is given by the following relation:
\begin{align}
\label{dynamicmomentumharmonic}
&\frac{d}{dt} \omega_{N+1} \left(\hat{A} \, \alpha^{t}_{N+1}\left(\hat{P}_{cm}^{N+1} \right) \hat{B}\right) = \\
&\omega_{N+1}\left(\hat{A} \left[\sum_{j = - \frac{N}{2}}^{\frac{N}{2}} \hat{p}_j, \sum_{k = - \frac{N}{2}}^{\frac{N}{2}} \frac{1}{2} m_{k}^{N+1} \nu^2 \hat{x}_{k}^{2}\right]\hat{B}\right) \nonumber \\
& \to - M \nu^2 \omega(\hat{X}_{cm}). \nonumber
\end{align}
We can easily recognize the classical dynamics of point particles given by Poisson brackets in the Hamiltonian formalism, where an effective harmonic potential $\frac{1}{2} M \nu^2 X_{cm}^{2}$ is applied to the macroscopic system.
\end{remark}

\section{Final remarks and conclusions}
\label{sec:end}
In this paper we proved how classical mechanics is achieved as quantum mechanics of a many-body system with infinite number of particles. We proved that the center of mass behaves as a classical system, and quantum mechanical effects such as quantum superposition cannot occurs at the macroscopic level. This can be achieved by quite general assumptions on the nature of the system and on the interaction among the particles. 

Despite the fact that we proved this for quite general states $\omega$, in general, from physics points of view one of the main interesting case is where the state $\omega$ is a ground state. It has been proved that the macroscopic state $\omega$ obtained as limit of the ground states $\omega_N$ is also a ground state of the macroscopic system, and that the system always admits ground states which are also pure states, and all the other states with minimal energy can be written as convex sum of the pure ground states (see~\citep{bratteli2012operator}, propositions 5.3.23, 5.3.24 and 5.3.37 for the proofs). It is, however, not possible to say if $\omega$ is also pure by looking only at the center of mass properties, but the overall state should be considered (i.e. including also the relative coordinates). For example, if all the state $\omega_N$ are invariant with respect to a certain symmetry group but the macroscopic state $\omega$ is not pure and the pure states are not invariant, then there is a spontaneous symmetry breaking. However, this goes beyond the goal of this paper, since here we just focus on the center of mass properties of the macroscopic state, and we just say that any pure ground state has center of mass properties localized in both position and momentum.

The interaction among particles here assumed need to be attractive and strong enough, in order to have a constraint in the volume of space the particles can occupy around their center of mass, and it needs to be effectively short-ranges, i.e. at any fixed time $t>0$ the correlations spreads at finite speed and thus preventing any infinite-range correlation.

The conditions here described are sufficient to achieve the quantum-to-classical transition, but it might be possible to relax at least some of them. Moreover, more studies have to be done to check how quantum-to-classical transition might occur in a relativistic system. We can, however, say that the result achieved here allows to perform an important step towards a unique theory describing quantum mechanics, classical non-relativistic mechanics and general relativity.

As shown in this paper, it is not necessary to modify quantum mechanics to describe the components of matter, nor is necessary to play with the interpretations of quantum mechanics. Finally, no external environment is necessary to achieve classical mechanics. Classical mechanics, where the state of objects are described as points in phase space and where the dynamics is described by Poisson brackets, and where observables are continuous functions on phase space, is a simple consequence of the macroscopicity of classical matter when compared to the microscopic components. When we measure properties of the global, macroscopic objects, we lose any information on the microscopic particles composing the matter and classical mechanics works. On the other side, when we want to observe microscopic systems, composed by a finite number of particles, quantum mechanical properties of particles appear and cannot be neglected.
\section*{Acknowledgements}
The author want to thank Dr. M. Toros and Dr. G. Gasbarri for continuous discussions, helpful for finalizing this paper.
\bibliography{bibarticle.bib}
\end{document}